\def\AaA{{A\&A}}
\def\AJ{{\em Astr.~J.}}

\def\MN{{MNRAS}}

\def\spose#1{\hbox to 0pt{#1\hss}}
\def\approxlt{\mathrel{\spose{\lower 3pt\hbox{$\sim$}}
	\raise 2.0pt\hbox{$$<$$}}}
\def\approxgt{\mathrel{\spose{\lower 3pt\hbox{$\sim$}}
	\raise 2.0pt\hbox{$>$}}}

\def\multleft#1{\hbox to size{\vbox {\halign {\lft{##}\cr #1}}\hfill}\par}
\def\multright#1{\hbox to size{\vbox {\halign {\rt{##}\cr #1}}\hfill}\par}

\def\today{\ifcase\month\or January\or February\or March\or April\or May\or
      June\or July\or August\or September\or October\or November\or December\fi
      \space\number\day, \number\year}
\def\$<${\thinspace}
\def\boxit#1{\vbox{\hrule\hbox{\vrule\kern3pt\vbox{\kern3pt
          #1 \kern3pt}\kern3pt\vrule}\hrule}}

\documentstyle[psfig,times]{mn}
\begin{document}
\hsize=6truein

\title[The optical variability of the Seyfert 1 galaxy IRAS 13224-3809]
{The optical variability of the narrow line Seyfert 1 galaxy IRAS
13224-3809}
\author[Young et al.]
{\parbox[]{6.in}{A.J. Young$^1$, C.S. Crawford$^1$, A.C. Fabian$^1$,
W.N. Brandt$^2$ and P.T. O'Brien$^3$ \\
\footnotesize
$^1$\emph{Institute of Astronomy, Madingley Road, Cambridge CB3 0HA}
\\ $^2$\emph{Department of Astronomy, The Pennsylvania State
University, University Park, PA 16802, USA} \\ $^3$\emph{Department of
Physics \& Astronomy, University of Leicester, University Road,
Leicester LE1 7RH} \\}}
\maketitle

\begin{abstract}
We report on a short optical monitoring programme of the narrow-line
Seyfert 1 Galaxy IRAS 13224-3809. Previous X-ray observations of this
object have shown persistent giant variability. The degree of
variability at other wavelengths may then be used to constrain the
conditions and emission processes within the nucleus. Optical
variability is expected if the electron population responsible for the
soft X-ray emission is changing rapidly and Compton-upscattering
infrared photons in the nucleus, or if the mechanism responsible for
X-ray emission causes all the emission processes to vary together. We
find that there is no significant optical variability with a firm
upper limit of 2 per cent and conclude that the primary soft X-ray
emission region produces little of the observed optical emission. The
X-ray and optical emission regions must be physically distinct and any
reprocessing of X-rays into the optical waveband occurs some distance
from the nucleus. The lack of optical variability indicates that the
energy density of infrared radiation in the nucleus is at most equal
to that of the ultraviolet radiation since little is upscattered into
the optical waveband. The extremely large X-ray variability of IRAS
13224-3809 may be explained by relativistic boosting of more modest
variations. Although such boosting enhances X-ray variability over
optical variability, this only partially explains the lack of optical
variability.
\end{abstract}


\section{Introduction}

X-ray observations by Boller et al (1997) have shown persistent giant
variability on short timescales, with an amplitude of variability far
in excess of that seen in a typical broad line Seyfert 1 galaxy.  Such
extreme variability is determined by the emission and variability
mechanisms in the nucleus. The degree of variability at other
wavelengths may be used to constrain the conditions and
mechanisms. Optical variability will occur if, for example, the X-ray
emitting electron population is rapidly changing and Compton
scattering infrared radiation in the nucleus, or if the mechanism
responsible for the X-ray variations causes all the emission processes
to vary together. It has also been suggested (Boller et al 1997) that
such large amplitude X-ray variability may be due to relativistic
boosting, the degree of which is spectrally dependent.  We have
therefore observed IRAS 13224-3809 on three consecutive nights in
order to determine the level of optical variability in the source.

Previous studies of the narrow line Seyfert 1 galaxy IRAS 13224-3809
have shown it to be variable both in the UV and X-ray wavebands. Long
timescale Ly$\alpha$ line variability has been observed, with the line
profile and flux changing between three IUE observations over four
months (Mas-Hesse et al. 1994). UV continuum variability has been
observed in 11 IUE observations spanning three years, with the flux
varying by 24 per cent (Rodr\'{\i}guez-Pascual 1997).

\section{Optical observations}

IRAS 13224-3809 ($\alpha_{2000}=13^{\rm h}25^{\rm m}19^{\rm s}$,
$\delta_{2000}=-38^{\rm h}24^{\rm m}53^{\rm s}$) was observed in
\emph{B}, \emph{V}, \emph{R} and 
\emph{I} on three consecutive nights from 1997 March 13 to 16 using
the blue sensitive TEK4 CCD at the Cassegrain focus of the 1m Jacobus
Kapteyn Telescope (JKT). The $5.6\times5.6$ arcmin CCD field contained
the galaxy and a number of foreground stars (see Fig.~1) allowing us
to perform relative photometry. The images were bias subtracted and
flat fielded using sky flats. Flux calibration was performed using the
standard stars PG0942-029B, C and D of Landolt (1992). IRAS 13224-3809
has a foreground star 7.6 arcsec from the optical nucleus (10.7 arcsec
from the X-ray centroid position (Boller et al. 1997)), and the
aperture used to collect the light from the nucleus was chosen to have
a diameter of 6.6 arcsec. This was found to collect most of the light
from the nucleus whilst minimising that from the surrounding galaxy
and the foreground star. Repeating our analysis with different sized
apertures did not significantly affect our results. The same size
aperture was used to obtain the flux from a comparison star in the
field. Fig.~2 shows the radial profile of pixel values from the centre
of IRAS 13224-3809. The scatter in pixel value is due to the different
contributions to the radial profile from different azimuths. The
contribution from the nucleus, underlying galaxy, the nearby star and
background may be seen.

\begin{figure}
\centerline{\psfig{figure=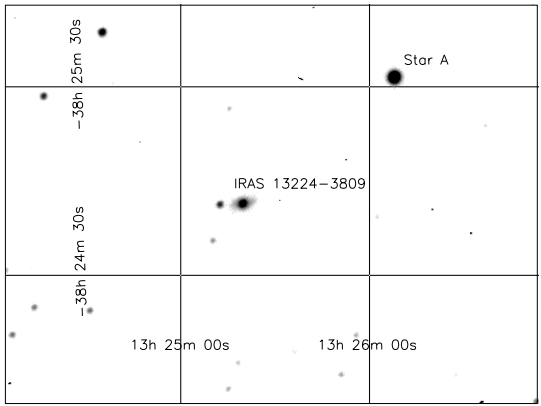,width=0.4\textwidth,angle=0}}
\caption{V-band image of IRAS 13224-3809 ($\alpha_{2000}=13^{\rm
h}25^{\rm m}19^{\rm s}$, $\delta_{2000}=-38^{\rm h}24^{\rm m}53^{\rm
s}$) and a comparison star A.}
\end{figure}

\begin{figure}
\centerline{\psfig{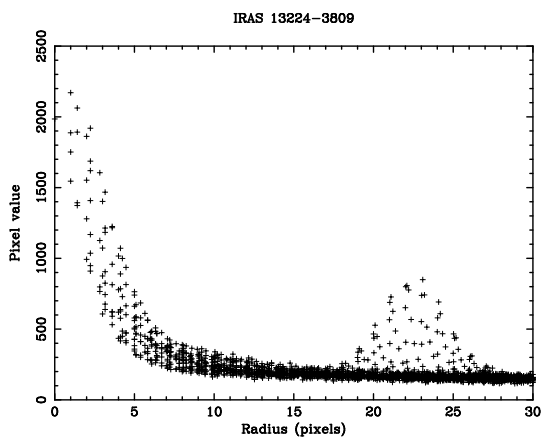}}
\caption{Radial profile of IRAS 13224-3809. The effect of the point
source close to the nucleus can be seen at a distance of about 23
pixels from the centre. The scatter in pixel value is due to the
contribution from different azimuths.}
\end{figure}

We follow the method of Done et al (1990) in which the variations seen
in a comparison star due to small-scale rapid changes in seeing are
used as a scalable template to model the variability one would expect
to see in the nucleus as a results of those seeing changes. During
each night the zenith distance of IRAS 13224-3809 varied between
67--72$^\circ$ and over such a small range we expect the extinction to
remain almost constant.  Since the images were taken through a number
of different filters we grouped the images by night and by filter and
then normalised them so that they have the same mean. This allows us
to compare images taken with different filters. Fig.~3 shows the
stellar and galactic residuals about the mean for a typical night. We
would expect the slope of the best fitting line to be 1 if the seeing
had an equal effect on both the galaxy and the comparison star. The
actual fit has a slope of 1.10 indicating that seeing has a larger
effect on the flux measured for the galaxy than the star, assuming
both are not variable. The stellar residuals are then used as a
scalable template for the variations of the galaxy about the mean flux
of the galaxy. The normalisation of this template is chosen to
minimise the difference between the predicted and observed galactic
fluxes. Fig.~4 shows the template light curve, created using the
scaled stellar residuals, and the actual light curve.  There is very
good agreement between the two.

Fig.~5 shows the light curve for the three nights using data from all
filters, with $1\sigma$ Poisson error bars. The light curve is
consistent with less than 1 per cent variability during each
night. One may expect the amplitude of variability to be wavelength
dependent, but we are unable to detect significant variability in the
data from any one filter.  The light curve for each night is
normalised to the mean for that night although there is less than 1
per cent variability between nights. The large breaks in the light
curve are due to high winds at the telecsope preventing observing.

\begin{figure}
\centerline{\psfig{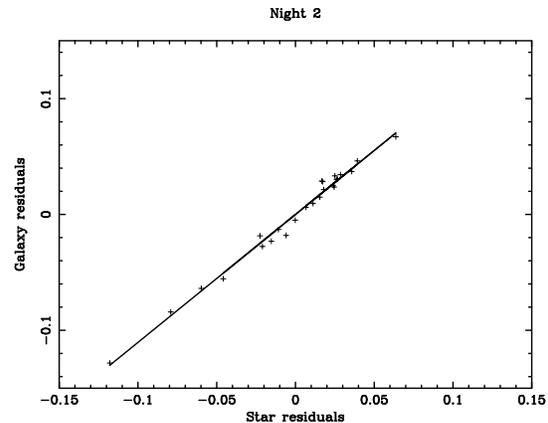}}
\caption{The residuals about the mean for the nucleus and star
for the second night's data. The best fitting straight line has a
slope of 1.10. This indicates that (were both sources to be constant)
the seeing has a larger effect on the galaxy than the star.}
\end{figure}

\begin{figure}
\centerline{\psfig{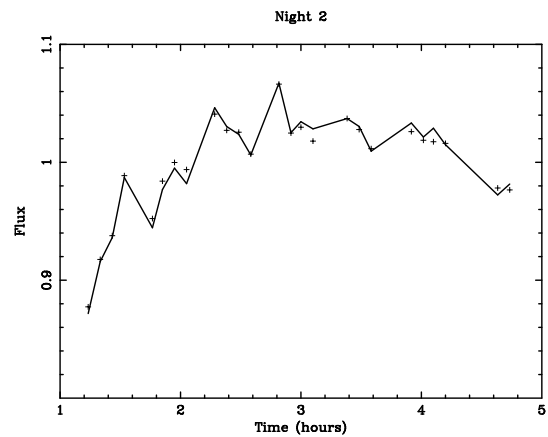}}
\caption{Flux of the nucleus predicted using the scaled stellar residuals
(solid line) and the actual observed flux of the galaxy.}
\end{figure}

\section{Discussion}

\subsection{Optical}

\begin{figure}
\centerline{\psfig{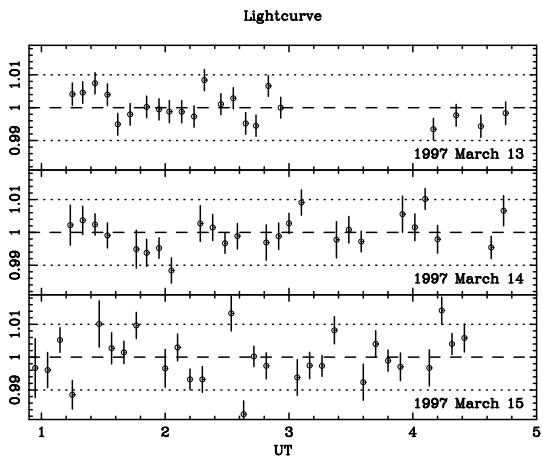}}
\caption{The light curve for the three nights. It is consistent with the
source having varied by less than 1 per cent during the course of our
observations.}
\end{figure}

The optical light curve of IRAS 13224-3809 is consistent with the
short time-scale variability of the source being less than 1 per
cent. We are unable to comment on the wavelength dependence of any
optical variability. Our aperture, however, also collects flux from
the underlying galaxy which dilutes the signal from the nucleus.  As
may be seen from Fig.~1 the three main components of the flux received
within a radius of 10 pixels, namely the contribution of the
background, the underlying galaxy and the nucleus may be separated. If
the nucleus is assumed to be a Gaussian point source and we (over-)
estimate the contribution from the galaxy then the variability of the
nucleus may be constrained. It is found that a 1 per cent variation of
the entire source may correspond to a $\sim 2$ per cent variation in
the nucleus.

The continuum optical spectrum of IRAS 13224-3809 (Boller et al 1993)
has a photon index of approximately $\Gamma_{\rm o}\sim 0$ across the
optical waveband. The V band magnitude of the nucleus is 15.2,
corresponding to a $\nu F_\nu$ flux in the V band of
$1.5\times10^{-11}$ erg cm$^{-2}$ s$^{-1}$ at 5500\AA.

\subsection{X-ray}

Simultaneous X-ray data are not available but if the 30 day ROSAT
light curve of Boller et al (1997) is assumed to be typical of IRAS
13224-3809 it is extremely unlikely that the X-ray flux did not at
least double during the period of our observations. This light curve
is shown in Fig.~6 and there is no set of three consecutive days
during which the X-ray flux did not change considerably. The only
period of relative `quiescence' is between 8--10 days, and even during
this the X-ray flux at least doubles. A more recent HRI monitoring
campaign has confirmed the continued existence of such extreme
variability (private communication).

\begin{figure}
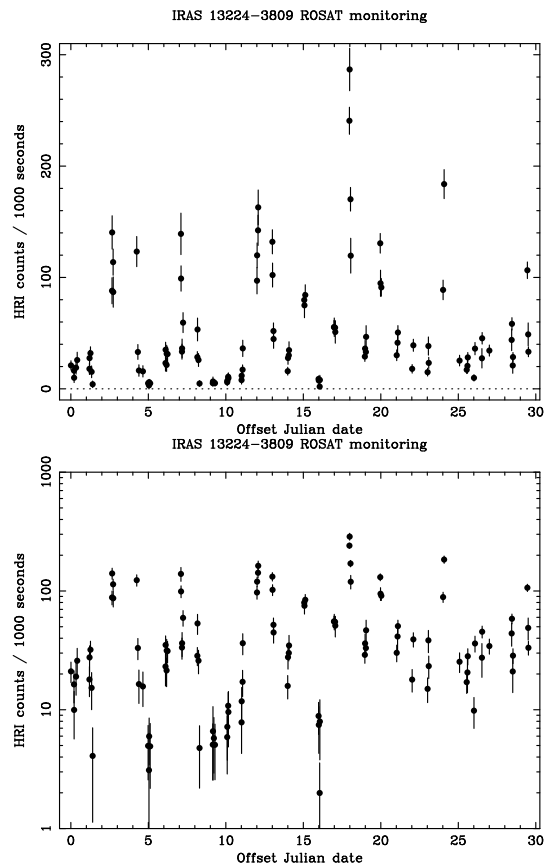

\centerline{\psfig{figure=hri_lc_lin.ps,width=0.4\textwidth,angle=270}}
\centerline{\psfig{figure=hri_lc_log.ps,width=0.4\textwidth,angle=270}}
\caption{The ROSAT 30-day HRI light curve of IRAS 13224-3809 (Boller
et al 1997) in linear scale (upper panel) and logarithmic scale (lower
panel). Note that the source varies cosiderably even during apparently
quiet behaviour in the upper panel, e.g. days 8--10.}
\end{figure}

The X-ray photon index of IRAS 13224-3809 was observed to be very
high, $\Gamma_{\rm x}\sim 4$, and the average HRI count rate was
$4.7\times10^{-3}$ count s$^{-1}$ (Boller et al 1997) corresponding to
a $\nu F_\nu$ flux of $2.9\times10^{-13}$ erg cm$^{-2}$ s$^{-1}$ at
1.25 keV.

\subsection{The nuclear X-ray and optical emitting regions}

Optical emission can originate from the soft X-ray emission region, or
very close to it, by several means. It may be emitted by the same
electrons that produce the X-rays, by cyclotron or synchrotron
emission or Comptonization (perhaps of photons produced by the
cyclo-synchrotron process; see e.g. Di Matteo, Celotti \& Fabian
1997). Comptonization  is however unlikely to make a large
contribution to the observed flux since the steep X-ray spectrum
implies a low Compton $y$-parameter which gives a similarly steep
optical spectrum.  Optical emission  may also originate from
reprocessing of the X-ray flux, and therefore would have a thermal
spectrum (probably from a range of temperatures). Finally it may be
triggered by whatever mechanism  causes the X-ray variability but not
share the same emitting electrons as the X-rays.

The X-ray flux is expected to have at least varied by a factor of two
during our observations. Let us assume that the X-ray flux doubled at
some point and consider the implications of the lack of optical
variability.

A lack of variability in the optical band can be explained in several
ways.  It may simply be due to most, more than 98 per cent, of the
optical emission emerges from a region unconnected with that producing
the soft X-ray flux. Note however that the average soft X-ray flux
(below 1 keV) is comparable to the optical flux and the peak soft
X-ray flux may be significantly greater.  It could also be due to  the
presence of many very dense clouds in the nucleus which free-free
absorb the intrinsic optical flux (Celotti, Rees \& Fabian
1994). Alternatively, the spectrum of the X-ray source may be such
that relativistic effects preferentially enhance the variability of
the X-ray emission over the optical emission.

The exceptionally large and rapid X-ray variability of IRAS 13224-3809
may be explained if the X-ray emission is relativistically boosted
(Boller et al 1997). For a source moving at a fraction $\beta$ of the
speed of light inclined at an angle $i$ to the observer the Doppler
parameter is given by $\delta=[\gamma(1-\beta\sin i)]^{-1}$, where
$\gamma=(1-\beta^2)^{-\frac{1}{2}}$. If the source is a power law of
photon index $\Gamma$ this gives rise to a boost in the amplitude of
variability by a factor of $\delta^{3+\Gamma}$. Fig.~7 shows the X-ray
and optical boost factors that may be produced as a function of radius
for different continuum photon indices and accretion disc
inclinations. It is possible for the X-ray boost factor to be many
times greater than the optical boost factor. (The ratio of the X-ray
to optical boost factor is $\sim\delta^{\Gamma_{\rm x}-\Gamma_{\rm
o}}$). The fraction of the optical emission that may be produced by,
or whose variability is tied to, the soft X-ray emitting regions may
then be constrained. The maximum fraction $0.02\delta^{\Gamma_{\rm
x}-\Gamma_{\rm o}}$ of the optical emission that is due to the X-ray
source is shown in Fig.~8 assuming values of $\Gamma_{\rm x}=4$ and
$\Gamma_{\rm o}=0$.  For reasonable values of the inclination and
emission radius it is unlikely that the X-ray source is responsible
for more than about 10 per cent of the optical flux.

The observed photon index of the optical continuum may differ from
that of the optical emission associated with the X-ray source. The
lowest possible $\Gamma_{\rm o}$ is $-1$ from the Rayleigh-Jeans part
of the blackbody spectrum, and it is possible that $\Gamma_{\rm x}$
exceeds 4. It remains unlikely, however, that more than about 20 per
cent of the optical emission is closely associated with the soft X-ray
emission. Differential relativistic effects between the X-ray and
optical bands can therefore be up to a factor of 10 but not completely
explain the lack of optical variability.

The hot electron population responsible for the X-ray emission can
inverse-Compton-scatter lower energy infrared photons into the optical
waveband and similarly ultraviolet photons into the X-ray
waveband. The observed lack of variability can be used to place a
limit on the ratio of the energy density of infrared to ultraviolet
radiation in the nucleus.  The observed ratio of 1.25 keV X-ray to
optical luminosity, $L_{\rm x}/L_{\rm o}\sim10^{-2}$, and the lack of
optical variability suggests that the variable X-rays have $L_{\rm x
(var)}/L_{\rm o (var)}$ less than or comparable with 1, so, in the
nucleus, the energy density of infrared radiation $\varepsilon_{\rm
IR}$ is at most that of the ultraviolet radiation $\varepsilon_{\rm
UV}$. If we assume the source to have an X-ray photon index
$\Gamma_{\rm x}=4$, the 1.25 keV flux implies the 0.1 keV $\nu F_\nu$
flux will be $4.5\times10^{-11}$ erg cm$^{-2}$ s$^{-1}$ allowing the
stronger constraint $\varepsilon_{\rm IR}<0.01\times\varepsilon_{\rm
UV}$

\begin{figure}
\centerline{\psfig{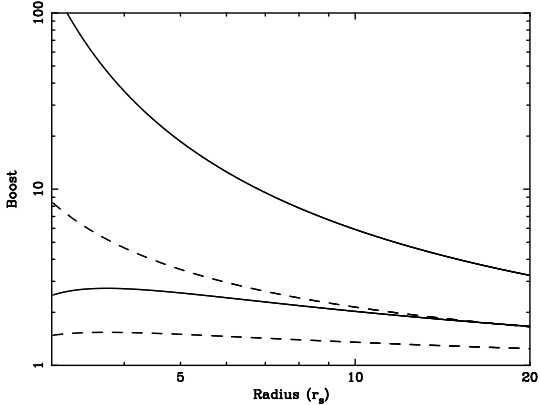}}
\caption{X-ray ($\Gamma_{\rm x}=4$, solid) and optical ($\Gamma_{\rm o}=0$, dashed)
boost factors as a function of radius. Curves are shown for two disc
inclinations, $30^\circ$ (lower curves) and $90^\circ$ (upper
curves). Velocities in the disc were calculated using a Newtonian
pseudo-potential.}
\end{figure}

\begin{figure}
\centerline{\psfig{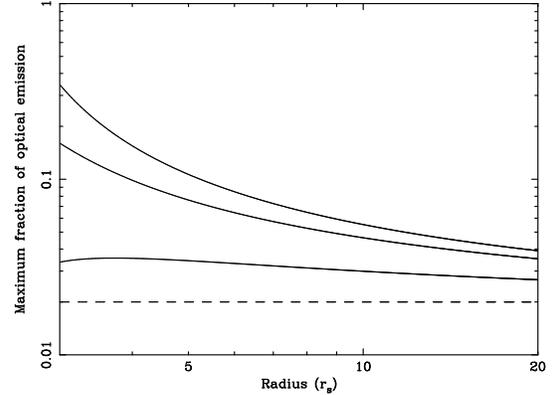}}
\caption{For a given radius and inclination the maximum fraction of the
optical emission that may be produced by the X-ray source is
calculated. The solid curves are for inclinations of $30^\circ$,
$60^\circ$ and $90^\circ$ (increasing upwards). This assumes that the
X-ray flux is seen to double and the optical flux increases by at most
2 per cent. The dotted curve shows the limit on the optical fraction
if the variability is not due to relativistic effects.}
\end{figure}

\subsection{Location of any reprocessing material}

The lack of optical variability implies that very little of the
primary soft X-ray flux is reprocessed into the optical waveband
within approximately a light day of the source. This may be due to the
inner disc where the X-rays originate being hot and small, radiating
mostly in the EUV, and to a lack of optically thick material oriented
to reprocess these X-rays at larger radii. For a $3\times10^7M_\odot$
black hole two light days corresponds to roughly 1000 Schwarzschild
radii.

\section{Conclusion}

The Narrow Line Seyfert 1 galaxy IRAS 13224-3809 shows persistent
giant amplitude X-ray variability yet almost no optical variability
exceeding 2 per cent. Such an extreme difference in variability
between the two wavebands suggests the X-ray emitting regions do not
produce any optical variability. This is expected if the electron
populations responsible for X-ray and optical emission are physically
distinct and the spectrum of the X-ray source intrinsically produces
almost no optical emission. The energy density of infrared radiation
in the nucleus is at most equal that of the ultraviolet radiation
since the X-ray emitting electrons are not responsible for significant
inverse Compton scattering into the optical waveband. If relativistic
boosting occurs then the conclusions are weakened somewhat with at
most about 20 per cent of the optical emission associated with the
X-ray source. Any significant reprocessing of primary soft X-rays into
the optical waveband occurs on scales greater than $1000R_{\rm s}$ The
lack of optical variability is consistent with observations of less
extremely X-ray variable Seyfert 1 galaxies such as NGC 4051 (Done et
al 1990).

\section{Acknowledgements}

AJY thanks PPARC for support. ACF and CSC thank the Royal Society.

\end{document}